\documentclass[%
 reprint,
 amsmath,amssymb,
 aps,
onecolumn]{revtex4-2}

\usepackage{graphicx}% Include figure files
\usepackage{dcolumn}% Align table columns on decimal point
\usepackage{bm}% bold math
\usepackage{longtable}
\usepackage{bm}
\usepackage{relsize}
\usepackage{amsfonts}
\usepackage{amsmath}
\usepackage{amssymb,epsf}
\usepackage{latexsym}
\usepackage{graphicx,epsfig}
\usepackage{amssymb}
\usepackage{float}
\usepackage{subfigure}
\usepackage[export]{adjustbox} %for valign in fig
\usepackage{epstopdf}
\usepackage[colorlinks=true,citecolor=blue,linkcolor=blue,urlcolor=black]{hyperref}
\usepackage{dcolumn}
\usepackage{psfrag}
\usepackage{wrapfig}
\usepackage{makeidx}
\usepackage{epsf}
\usepackage{color}
\usepackage{multirow}
\usepackage{mathtools}
\usepackage{soul}
\begin{document}

\preprint{APS/123-QED}

\title{Dynamical reconstruction of the $\Lambda$CDM model in the scalar-tensor representation of $f\left(Q,T\right)$ gravity}% Force line breaks with \\

\author{Adam Z. Kaczmarek$^1$}
\email{adamzenonkaczmarek@gmail.com}
\author{João Luís Rosa$^{2,3}$}
\email{joaoluis92@gmail.com}
\author{Dominik Szcz{\c{e}}{\'s}niak$^1$}
\email{d.szczesniak@ujd.edu.pl}
\affiliation{$^1$Institute of Physics, Jan D{\l}ugosz University in Cz{\c{e}}stochowa, 13/15 Armii Krajowej Ave., 42200 Cz{\c{e}}stochowa, Poland}
\affiliation{$^2$Institute of Physics, University of Tartu, W. Ostwaldi 1, 50411 Tartu, Estonia}
%\affiliation{$^{2,3}$Institute of Physics, Jan D{\l}ugosz University in Cz{\c{e}}stochowa, 13/15 Armii Krajowej Ave., 42200 Cz{\c{e}}stochowa, Poland}
\affiliation{$^3$Institute of Theoretical Physics and Astrophysics, University of Gdańsk, Jana Ba{\.z}yńskiego 8, 80309, Gda{\'n}sk, Poland}
\date{\today}

\begin{abstract}
Motivated by the growing interest in the nonmetricity-matter couplings, we develop the scalar-tensor formulation of recently introduced $f(Q,T)$ gravity, where $Q$ is the nonmetricity and $T$ is the trace of the energy-momentum tensor. The main properties of the scalar-tensor formalism for the Friedmann-Lema{\^ i}tre-Robertson-Walker (FLRW) Universe are discussed, and we introduce an appropriate set of dynamical variables to analyze the cosmic evolution of the scalar-tensor $f(Q,T)$ cosmology as a dynamical system. By considering two distinct cosmic fluids, namely matter and radiation, we have demonstrated that the cosmological phase space exhibits the typical curvature-dominated, radiation-dominated, matter-dominated, and exponentially accelerated fixed points. Furthermore, under an appropriate set of initial conditions compatible with the current observations from the Planck satellite, our analysis shows that the scalar-tensor $f(Q,T)$ successfully yields models indistinguishable from the $\Lambda$CDM cosmology and compatible with the weak-field solar system dynamics, without the inclusion of a cosmological constant $\Lambda$. Thus, the theory introduced herein may be regarded as a suitable candidate to describe the cosmological dynamics of the Universe.
\end{abstract}

%\keywords{Suggested keywords}%Use showkeys class option if keyword
                              %display desired
\maketitle

%\tableofcontents

\section{Introduction}
The observation that the Universe is expanding at an accelerated rate was a scientific breakthrough that revolutionised the way in which we perceive the Cosmos. The current concordance model of the Universe, proposed to explain this accelerated expansion, is associated with the $\Lambda$CDM paradigm and the introduction of an unobserved energy component known as the dark energy (in form of the cosmological constant $\Lambda$). To overcome this limitation, researchers are exploring alternative approaches not only in particle physics but also in modifying Einstein's General Relativity (GR) \cite{clifton2012,joyce2016dark,nojrii2017}.
Several modifications of GR have been proposed, following the path set by Eddington and developed later by Buchdachl and Horndeski \cite{buchdachl1970,Horndeski1974}. These proposals have shown promise not only in explaining the cosmic expansion, but also in resolving other outstanding problems in modern cosmology. Moreover, the methods of dynamical systems were shown to be particularly useful in the analysis of the cosmological evolution in modified theories of gravity \cite{bahamonde2018dynamical}. As such, modifications of GR are also often considered as the more economical way in solving particular issues in comparison to the conventional general theory of relativity \cite{sebastiani2017,nojrii2017}. 

One of the novel schemes to modify gravity focuses on the inclusion of the nonminimal coupling between matter and geometry, allowing the Lagrangian to depend on both curvature and matter, usually represented as the trace of energy-momentum tensor ($T$) \cite{harko2011,harko2014b}. The main examples of such theories are $f(R,T)$ gravity \cite{harko2011}, its Gauss-Bonnet counterpart ($f(\mathcal{G},T)$) gravity \cite{sharif2016}, or even the energy-momentum squared gravity \cite{cipriano2024}.  In this manner, it is possible to describe both (interacting) dark energy and dark matter, where the former is related to the nongeodesic motion of particles \cite{harko2014b}. The $T-$dependence of these theories is motivated by the requirement of the conformal invariance of the physical laws, the concept introduced by Weyl \cite{weyl1918,lobo2022curvature}. An interesting feature of these theories is their possible interpretation as an effective description of specific quantum gravity phenomena \cite{yang2016effects}. Another advantage of geometry-matter couplings is associated with the interesting phenomena that may be described in that framework, e.g. particle production induced by gravity \cite{cipriano2024gravitationally,pinto2022gravitationally} and non-exotic wormholes \cite{Rosa:2022osy,Rosa:2023guo}. Since equations of motion in these extended theories, with additional scalar degrees of freedom in comparison to GR, usually present a high degree of complexity, it is not unusual to recur to alternative scalar-tensor representations in an attempt to simplify the task of solving them \cite{Rosa:2021teg}.

Interestingly, the curvature and geometry defined by the Levi-Civita connection are not the only possible options to describe gravity. Alternative approaches are built upon the idea of telleparallelism, and belong to a broader set of metric-affine theories of gravity \cite{gronwald1997metric,ortin2004gravity,jimenez2022metric,bahamonde2023teleparallel}. Theories based on these approaches may be valid at the Solar System level and also provide an explanation for the properties of the Universe at large cosmological scales \cite{bahamonde2023teleparallel,arora2021constraining}. As such, perhaps one of the most influential extension of the Riemannian geometry can be attributed to Cartan. The theory is known as the Einstein-Cartan gravity, where another set of equations couples torsion to the matter \cite{Cartan1923,cartan1924,Cartan1925}. This theory described gravity through both curvature and torsion. This inspired further searches in the field of telleparalel gravity, where spacetime is characterized by a non-vanishing torsion, but without curvature \cite{bahamonde2023teleparallel,blixt2021review}. Besides curvature and torsion, gravity can also be attributed to the nonmetricity ($Q$), i. e. the change of the vector length under parallel transport \cite{jimenez2018coincident,bahamonde2023teleparallel}. In fact, those three approaches are equivalent to GR, often referred to as the trinity of gravity \cite{beltran2019geometrical}. 

The nonmetricity gravity (also known as STEGR - symmetric teleparallel equivalent of general relativity) has been quickly extended to the broad family of $f(Q)$ gravity models. This approach opened new opportunities in studying alternatives to GR, similarly to the impact of $f(R)$ theories for Riemannian gravity \cite{capozziello2007,clifton2012,nojiri2017}. Notably, while the equations of motion in the $f(R)$ gravity are of higher order, the $f(Q)$ theories yield second-order field equations \cite{jimenez2018,jarv2018,barros2020,hu2022}. Although relatively new, the $f(Q)$ gravity has already produced interesting results, challenging the $\Lambda$CDM model and providing consistent alternatives to describe the Universe \cite{beltran2019,atayde2021,anagnostopoulos2021,capozziello2022model}. Hence, models built from nonmetricity have been explored from multiple perspectives, including astrophysics and black holes \cite{lazkoz2019,bahamonde2022,calza2022,errehymy2022,mandal2022,calza2023}, as well as cosmology \cite{bajardi2020,solanki2021,d2022black,junior2023,paliathanasis2023dynamical,capozziello2023role,capozziello2024gravitational}. The framework of symmetric teleparallelism has been extended through the addition of different degrees of freedom \cite{jarv2018,kaczmarek2024npb}. Furthermore, Harko and colleagues introduced a coupling between nonmetricity and matter in the form $L\sim f_1(Q)+f_2(Q)L_M$ \cite{harko2018}, which was then further expanded to the $f(Q,T)$ gravity \cite{xu2019f} and gained a significant attention lately \cite{gadbail2021,shiravand2022cosmological,narawade2023,tayde2023existence,tayde2024conformally}. Moreover, it was recently argued by Nojrii and Odintsov \cite{NOJIRI2024} that a coupling between nonmetricity and matter may resolve one of the ambiguities related to the $f(Q)$ gravity. They have shown that one of the possible choice of gauge, called coincident gauge, is not compatible with the Friedmann universe, unless one couples $Q$ with matter. Thus, non-minimal couplings between nonmetricity and matter may be a possible way to avoid this kind of pathologies. 

It is noteworthy that theories with matter-geometry couplings such as $f(R,T)$ or $f(Q,T)$ can also be recast into a dynamically equivalent scalar-tensor theory, by a suitable reformulation of the action principle in terms of two scalar fields \cite{gonccalves2022cosmology,kaczmarek2024hybrid}. This approach was initially developed by Tamanini and Boehmer for the hybrid metric-Palatini gravity \cite{tamanini2013generalized}. In this manner, by the inclusion of additional degrees of freedom, it is possible to obtain an equivalent theory with equations of motion that are second order at most \cite{pinto2022gravitationally,gonccalves2022cosmology,kaczmarek2024hybrid,cipriano2024gravitationally}. This fact opens additional theoretical motivations for studying those approaches, as possible limits of the more fundamental theories, when different scalar fields couple with the distributions of matter and geometry. For instance, Damour and Polyakov explored theories where string-loop modifications of the low energy dilaton-matter coupling could fix the vacuum expectation value of a massless dilaton \cite{damour1994string}. Furthermore, recent developments in the Kaluza-Klein theory often introduce multiple scalar fields, such as the compacton \cite{damour1992}. That interplay motivates us to study the novel scalar-tensor reformulation of the $f(Q,T)$ gravity, with the goal of exploring its phenomenological aspects for cosmology.

Due to the complexity and richness of developed models, one of the most crucial aspects studied in the literature is their dynamical behaviour. By casting them in the form of the dynamical system, one gets the ability to provide qualitative and even quantitative information on the behaviour of dynamics in the studied cosmological model, in the form of the available states and constraints on the evolution \cite{mirza2016dynamical,bahamonde2018dynamical,rosa2024dynamical}. Various modified gravity models have been studied in that manner over the years, including $f(R)$ theory and some of its extensions \cite{shabani2013f,Carloni:2018yoz,Rosa:2019ejh,Rosa:2024pzo}, teleparallel $f(\mathcal{T})$ model \cite{ganiou2019cosmological,kadam2022teleparallel}, $f(Q)$ counterpart \cite{paliathanasis2023dynamical,bohmer2023dynamical} and importantly $f(R,T)$ gravity in the both standard and scalar-tensor formulation \cite{mirza2016dynamical,gonccalves2024dynamical}. It was observed that conservation of matter plays an important role in the dynamics within modified theories of gravity. Even though $f(Q,T)$ gravity has already been studied using the dynamical system approach \cite{narawade2023}, this analysis was heavily model dependent. Note, that one of the advantages of the scalar-tensor formulation is the general approach to the cosmological dynamics, as different $f(Q,T)$ models are characterised by different potentials, similarly to other scalar-tensor theories of gravity of that kind \cite{cipriano2024gravitationally}. Hence, scalar-tensor formulation allows for a more tractable analysis, which is carried out in this work.

The following work is organized as follows. In Sec. \ref{sec:theory}, the theoretical formulation of $f(Q,T)$ gravity is presented in both its original formulation and its equivalent scalar-tensor representation. In Sec. \ref{sec:dynsys}, we present the framework for the dynamical system analysis within the Friedmann-Lema{\^ i}tre-Robertson-Walker (FLRW) spacetime, we study the GR limit of the theory and identify the corresponding critical points in the phase space. Next, we numerically integrate the dynamical system in order to obtain dynamics consistent with the $\Lambda$CDM model. The manuscript concludes with a summary in Sec. \ref{sec:concl}.

\section{Theoretical formulation}\label{sec:theory}
\subsection{Original formulation of $f\left(Q,T\right)$ gravity}

The action that describes a generalized nonmetricity theory with nonminimal coupling between geometry and matter encoded in $Q$, trace of energy momentum tensor $T=T_\mu^{\;\mu}$ and matter Lagrangian $\mathcal{L}_m$ is given by \cite{xu2019f}:
\begin{align}
  S=\int \sqrt{-g}\Big[\frac{1}{16 \pi} f(Q, T) + \mathcal L_{m}\Big]d^4x,  
\label{eq1}
\end{align}
where we have adopted a set of geometrized units such that $G=c=1$, with $G$ being the gravitational constant and $c$ standing for the speed of light, $g$ is the determinant of the metric $g_{\mu\nu}$ written in the coordinate system $x^\mu$, and we have defined:
\begin{align}
Q\equiv -g^{\mu \nu}\big( L^{\alpha}_{\ \ \beta\mu}L^{\beta}_{\ \ \nu\alpha} - L^{\alpha}_{\ \ \beta\alpha} L^{\beta}_{\ \ \mu \nu}  \big),
\;\;\;\;
\text{and}
\;\;\;\;
L^{\alpha}_{\ \ \beta \gamma}\equiv -\frac{1}{2} g^{\alpha \lambda}\big(  \nabla_{\gamma}g_{\beta \lambda} + \nabla_{\beta}g_{\lambda\gamma}-\nabla_{\lambda}g_{\beta \gamma}  \big) ,
\end{align}
respectively. In the action given above, the non-minimal geometry-matter coupling in terms of the Ricci scalar $R$ present e.g. in the $f\left(R,T\right)$ theory of gravity is replaced by a nonmetricity-matter coupling \cite{xu2019f,arora2021constraining}. Generally speaking, the nonmetricity tensor characterizes a change in the length of a transported vector and is defined to be \cite{hehl1976general}:
\begin{align}
    Q_{\lambda\mu\nu}=-\nabla_\alpha g_{\mu\nu}=-\frac{\partial g_{\mu\nu}}{\partial x^\lambda}+g_{\nu\sigma}\hat{\Gamma}^\sigma_{\ \mu\lambda}+g_{\sigma\mu}\hat{\Gamma}^\sigma_{\  \nu\lambda},
\end{align}
for the (asymmetric) Weyl-Cartan connection $\hat{\Gamma}^{\lambda}_{\mu\nu}$. It is worth to remark that in the auxillary Weyl-Cartan geometrical description, the connection may be decomposed into three parts. Namely, the Christoffel symbols $\Gamma^\lambda _{\ \mu\nu}$ related to the usual Levi-Civita connection in GR, the contortion $C^\lambda_{\ \mu\nu}$ obtained from the torsion, and the disformation tensor $L^\lambda_{\ \mu\nu}$ associated with the nonmetricity. Thus, the connection can be written as:
\begin{align}
    \hat{\Gamma}^\lambda_{\  \mu\nu}=\Gamma^\lambda _{\ \mu\nu}+C^\lambda_{\ \mu\nu}+L^\lambda_{\ \mu\nu}.
\end{align}

\noindent Defining the two non-metricity traces $Q_{\alpha}$ and $\tilde{Q}_{\alpha}$, as well as a superpotential $P^{\alpha}_{\ \ \mu\nu}$ as follows:
\begin{align}
Q_{\alpha}\equiv Q_{\alpha\ \ \ \mu}^{\ \ \mu}, \ \ \ \ \tilde{Q}_{\alpha}\equiv Q^{\mu}_{\ \  \alpha \mu},
\end{align}
\begin{align}
 P^{\alpha}_{\ \ \mu\nu}\equiv \frac{1}{4}\bigg[ -Q^{\alpha}_{\ \ \mu \nu}+ 2Q^{\ \ \ \alpha}_{\big( \mu \ \ \ \nu \big)} + Q^{\alpha}g_{\mu \nu} - \tilde{Q}^{\alpha}g_{\mu\nu}\nonumber 
- \delta^{\alpha}_{\ \ ( \mu} Q _{\nu )} \bigg] = -\frac{1}{2}L^{\alpha}_{\ \ \mu\nu}+ \frac{1}{4}\big( Q^{\alpha} - \tilde{Q}^{\alpha}  \big) g_{\mu \nu} - \frac{1}{4} \delta^{\alpha}_{\ \ (\mu} Q_{\nu)}, 
\end{align}
where we have introduced the notation for index symmetrization $X_{(ab)=\frac{1}{2}\left(X_{ab}+X_{ba}\right)}$, the relationship between the quantities defined above holds:
\begin{align}
Q=-Q_{\alpha\mu\nu}P^{\alpha\mu\nu}=- \frac{1}{4}\big( -Q^{\alpha\nu\rho}Q_{\alpha\nu\rho}+ 2 Q^{\alpha\nu\rho} Q_{\rho\alpha\nu} 
- 2Q^{\rho}\tilde{Q}_{\rho} + Q^{\rho}Q_{\rho}  \big).
\end{align}

\noindent The stress-energy tensor $T_{\mu\nu}$ is defined in the usual way as:
\begin{align}
    T_{\mu\nu}=-\frac{2}{\sqrt{-g}}\frac{\delta (\sqrt{-g}\mathcal{L}_m)}{\delta g^{\mu\nu}},
\label{eq3}
\end{align}
whereas the auxiliary tensor $\Theta_{\mu\nu}$ is given as \cite{xu2019f,lobo2022curvature}:
\begin{align}\label{eq:deftheta}
        \Theta_{\mu\nu}=g^{\alpha\beta}\frac{\delta T_{\alpha\beta}}{\delta g_{\mu\nu}}.
\end{align}
Assuming that the matter Lagrangian $\mathcal{L}_m$ depends explicitly on the metric but not its derivatives, Eq. \eqref{eq:deftheta} takes the form:
\begin{align}
\Theta_{\mu\nu}=-2T_{\mu\nu}+g_{\mu\nu}\mathcal{L}_m-2g^{\alpha\beta}\frac{\partial^2 \mathcal{L}_m}{\partial g^{\mu\nu}\partial g^{\alpha\beta}}.
\end{align}

\noindent The field equations for the $f\left(Q,T\right)$ theory of gravity can be obtained through the variational method by taking a variation of Eq. \eqref{eq1} with respect to the metric tensor $g_{\mu\nu}$. This variation takes the form
\begin{align}\label{eq:field1}
    -\frac{2}{\sqrt{-g}}\nabla_{\alpha}(f_Q \sqrt{-g} P^{\alpha}_{\ \  \mu\nu})-\frac{1}{2}g_{\mu\nu} f(Q,T) + (T_{\mu\nu}+\Theta_{\mu\nu})f_T - f_Q(P_{\mu\alpha\beta}Q_{\nu}^{\ \ \alpha\beta}-P_{\alpha\beta\nu}Q^{\alpha\beta}_{\ \ \mu})=8\pi T_{\mu\nu}.
\end{align}
Additionally, the equations of motion for the connection can be obtained by taking a variation of Eq. \eqref{eq1} with respect to the curvature, with the associated Lagrange multiplier method, and together with vanishing curvature and torsion ($R_{\alpha\beta\mu\nu}=0$, $S_{\alpha\mu\nu}=0$), which leads to the result:
\begin{align}
    \nabla_\mu \nabla_\nu \Big( \sqrt{-g} f_Q P^{\mu\nu}_{\ \ \ \alpha}+4\pi H_{\alpha}^{\ \mu\nu}\Big)=0,\label{eq:connect}
\end{align}
where the hypermomentum $H_\alpha^{\ \mu\nu}$ is defined as \cite{xu2019f}:
\begin{align}
    H_{\alpha}^{\ \ \mu\nu}=\frac{\sqrt{-g}}{16\pi}f_T \frac{\delta T}{\delta \hat{\Gamma}^\alpha_{\ \ \mu\nu}}+\frac{\delta \sqrt{-g}\mathcal{L}_m}{\delta \hat{\Gamma}^\alpha_{\ \ \mu\nu}}.
\end{align}
Finally, the conservation equation for the $f(Q,T)$ gravity is obtained from a covariant derivative of Eq. \eqref{eq:field1} and it is given by:
\begin{align}
    \mathcal{D}_{\mu}\Big[ f_T (T^\mu_{\ \ \nu}+\Theta^\mu_{\ \ \nu})-8\pi T^\mu_{\ \nu}\Big]+\frac{8\pi}{\sqrt{-g}}\nabla_\alpha \nabla_\mu H_\nu^{\ \alpha\mu}=\frac{1}{\sqrt{-g}}Q_\mu \nabla_\alpha(f_Q \sqrt{-g}P^{\alpha\mu}_{\ \ \ \nu}) +\frac{1}{2}f_T \partial_\nu T,
\end{align}
where, for a $(1,1)$-form tensor, we have the following decomposition for the covariant derivatives of $V^\mu_{\ \nu}$: $\nabla_\mu V^\mu_{\ \nu}=\mathcal{D}_\mu V^{\mu}_{\ \nu}-\frac{1}{2}Q_\alpha V^\alpha_{\ \nu}-L^\alpha_{\ \mu\nu}V^\mu_{\ \alpha}$. 

\subsection{Scalar-tensor representation of $f(Q,T)$ gravity}

Similarly to some modified theories of gravity with additional scalar degrees of freedom, it is sometimes useful to introduce a dynamically equivalent scalar-tensor representation of the theory. This representation can be obtained by following the same approach as done in other theories, such as $f(R,T)$ gravity. We start by introducing two auxiliary fields $\alpha$ and $\beta$ and rewriting the action in Eq. \eqref{eq1} in terms of those fields as:
\begin{align}
    S=\frac{1}{16\pi}\int d^4x\sqrt{-g}\Big[ f(\alpha,\beta)+(Q-\alpha)f_{\alpha}+(T-\beta)f_{\beta}\Big] + \int d^4x\sqrt{-g} L_{m}.
\label{eq7}
\end{align}
The variation of the action in Eq. \eqref{eq7} with respect to the auxiliary fields $\alpha$ and $\beta$ leads to the two scalar equations that can be conveniently rewritten in the matrix form:
\begin{align}
 \textbf{M} \textbf{x}=   \begin{bmatrix}
f_{\alpha\alpha} & f_{\alpha\beta} \\
f_{\beta\alpha} & f_{\beta\beta} 
\end{bmatrix}  \begin{bmatrix}
Q-\alpha \\
T-\beta 
\end{bmatrix}=0,
\label{eq8}
\end{align}
where we have introduced the notation $f_{\alpha}\equiv\frac{\partial f(\alpha,\beta)}{\partial{\alpha}}$ and so on. Moreover, we  assume that function $f(\alpha,\beta)$ satisfies the Schwartz theorem, i.e. $f_{\alpha\beta}=f_{\beta\alpha}$. Then, the matrix form $\textbf{M} \textbf{x}=0$ has unique solution if and only if $\det \textbf{M}\neq 0$, which translates to $f_{\alpha\alpha}f_{\beta\beta}\neq f^2_{\alpha\beta}$ \cite{pinto2022gravitationally}. As a consequence, the unique solutions are $\alpha=Q$ and $\beta=T$, ensuring that Eq. \eqref{eq7} reduces to Eq. \eqref{eq1} and provides a direct equivalence between both of these formulations. 

Finally, introducing the following definitions for the scalar fields $\phi$ and $\psi$ as well as the scalar interaction potential $V(\phi,\psi)$:
\begin{align}
    \phi=\frac{\partial f}{\partial Q},\;\;\;\; \psi=\frac{\partial f}{\partial T},\;\;\;\; V(\phi,\psi)=\phi\alpha+\psi\beta -f(\alpha,\beta),
    \label{eq9}
\end{align}
the scalar-tensor representation of Eq. \eqref{eq7} reads:
\begin{align}
    S=\frac{1}{16\pi}\int d^4x\sqrt{-g}\Big[ \phi Q+\psi T - V(\phi,\psi)\Big] + \int d^4x\sqrt{-g} \mathcal L_{m}.
\label{eq10}
\end{align}
In this manner, we obtain the biscalar-nonmetric gravity with the non-trivial coupling between scalar field $\psi$ and matter via the $\psi T$ term. In other words, Eq. \eqref{eq10} can be regarded as a specific scalar-$Q$ gravity with an additional degree of freedom. The modified field equations in the scalar-tensor representation can now be obtained from a variation of Eq. \eqref{eq10} with respect to the metric tensor $g_{\mu\nu}$, which leads to:
\begin{align}
  -\frac{2}{\sqrt{-g}}\nabla_{\alpha}(\phi \sqrt{-g} P^{\alpha}_{\ \  \mu\nu})-\frac{1}{2}g_{\mu\nu} \big(\phi Q + \psi T - V(\phi,\psi)\big)+ \psi(T_{\mu\nu}+\Theta_{\mu\nu}) - \phi(P_{\mu\alpha\beta}Q_{\nu}^{\ \ \alpha\beta}-P_{\alpha\beta\nu}Q^{\alpha\beta}_{\ \ \mu})=8\pi T_{\mu\nu}
\label{eq11}
\end{align}%%%
while the equations of motion for $\phi$ and $\psi$ can be obtained from a variation of Eq. \eqref{eq10} with respect to these fields respectively, yielding: 
\begin{align}
    Q-\frac{\partial V}{\partial \phi}=0, \;\;\;\; T- \frac{\partial V}{\partial \psi}=0.
    \label{eq12}
\end{align}
Finally, the conservation equation in the scalar-tensor representation, obtained from a covariant derivative of Eq. \eqref{eq11}, takes the form:
\begin{align}
     \mathcal{D}_{\mu}\Big[ \psi (T^\mu_{\ \ \nu}+\Theta^\mu_{\ \ \nu})-8\pi T^\mu_{\ \nu}\Big]+\frac{8\pi}{\sqrt{-g}}\nabla_\alpha \nabla_\mu \mathcal{H}_\nu^{\ \alpha\mu}=\frac{1}{\sqrt{-g}}Q_\mu \nabla_\alpha(\phi\sqrt{-g}P^{\alpha\mu}_{\ \ \ \nu}) +\frac{1}{2}\psi \partial_\nu T,\label{eq:cons1}
\end{align}
where we have redefined the hypermomentum as:
\begin{align}
    \mathcal{H}_{\alpha}^{\ \ \mu\nu}=\frac{\sqrt{-g}}{16\pi}\psi \frac{\delta T}{\delta \hat{\Gamma}^\alpha_{\ \ \mu\nu}}+\frac{\delta \sqrt{-g}\mathcal{L}_m}{\delta \hat{\Gamma}^\alpha_{\ \ \mu\nu}}.
   \label{eq13}
\end{align}
As expected, the stress-energy tensor is not conserved for $\psi \neq 0$, which corresponds to the nonminimal coupling between nonmetricity and the trace of the stress-energy tensor $T$.
Using the equations of motion for the connection previously given in Eq. \eqref{eq:connect}, one can rewrite the conservation equation in Eq. \eqref{eq:cons1} in the more convenient form:
\begin{align}
    \mathcal{D}_{\mu}\Big[ \psi (T^\mu_{\ \ \nu}+\Theta^\mu_{\ \ \nu})-8\pi T^\mu_{\ \nu}\Big]-\frac{1}{2}\psi \partial_\nu T =\frac{1}{\sqrt{-g}}Q_\mu \nabla_{\alpha}(\sqrt{-g}\phi P^{\alpha\mu}_{\ \ \ \nu})+ \frac{2}{\sqrt{-g}}\nabla_\mu \nabla_{\alpha}(\sqrt{-g}\phi P^{\mu\alpha}_{\ \ \ \nu}).\label{eq:cons2}
\end{align}

\subsection{Choice of gauge and connection}

So far, our analysis remains general, because no assumptions on the shape of the connection $\hat{\Gamma}^\alpha_{\ \ \mu\nu}$ have been imposed. Since the scalar-tensor $f(Q,T)$ gravity is an extension of the symmetric telleparallel equivalent of general relativity (STEGR), both curvature and torsion are vanishing. Hence, nonmetricity alone drives the dynamics of gravity, leading to the following constraints:
\begin{align}
    R^{\alpha}_{\ \mu\nu\beta} =0,\;\;\; \mathcal{T}^{\alpha}_{\ \mu\nu}=0, \;\;\; \nabla_{\alpha} g_{\mu\nu}\neq 0.
\end{align}
Note that, since the vanishing of the curvature condition is imposed, the connection can be integrated as it means that the connection is the pure gauge field. Thus, the connection can be parametrized by an element  $\Lambda^{\alpha}_{\ \beta}$ of the linear $GL(4,\mathbb{R})$ group \cite{jimenez2018coincident,hu2022adm}:
\begin{align}
    \hat{\Gamma}^\mu_{\ \nu\alpha}=\big(\Lambda^{-1}\big)^\mu _\rho \partial_\alpha \Lambda^\rho_{\ \nu},
    \label{gaug}
\end{align}
leading to flat geometry. Additionally, the vanishing torsion condition guarantees the symmetry of the connection within the lower indices $\hat{\Gamma}^\mu_{\ \nu\alpha}=\hat{\Gamma}^\mu_{\ \alpha\nu}$ \cite{heisenberg2024review}. This leads to the additional constraint $\partial_{[\alpha}\Lambda^\rho_{\ \nu]}=0 $. Thus, using the parametrization $\Lambda ^\rho_{\ \nu}=\partial_\nu \xi^\rho$ by the vector field $\xi^\rho$, one can express the connection as \cite{jarv2018}:
\begin{align}
    \hat{\Gamma}^\alpha_{\ \mu\nu}=\frac{\partial x^\alpha}{\partial\xi^\beta}\partial_\mu \partial_\nu \xi^\beta,
\end{align}
In general, it is possible to choose a coordinate system such that the connection is removed. This can be achieved by aligning the vector field with the coordinate system, i.e., $\xi^\mu=x^\mu$. This choice is called a coincident gauge \cite{bahamonde2023teleparallel}. However, there are certain ambiguities with this choice of gauge. Specifically, not every coordinate system for FLRW cosmology is adapted to the requirement that all connection components vanish \cite{dimakis2022flrw,NOJIRI2024}. Thus, in what follows, we adopt a different choice of gauge, introduced by Nojiri and Odintsov in \cite{NOJIRI2024}.

\subsection{FLRW cosmology in the scalar-tensor representation}

With the purpose of the conducting a dynamical systems analysis of the scalar-tensor representation of $f(Q,T)$ in a cosmological context, we consider a Universe that is isotropic and homogeneous. Moreover, since the current observations seem to indicate that the Universe presents a flat geometry $(k=0)$, we introduce the flat Friedmann-Lemaitre-Robertson-Walker (FLRW) line element:
\begin{align}
    ds^2=-dt^2+a^2(t)\delta_{ij}\text{d}x^i\text{d}x^j,
\end{align}
for a scale factor $a(t)$ and the $x^i=\{x,y,z\}$ the spacial Cartesian coordinates. Additionally, we adopt the connection associated with the vector field $\xi^\mu=\big(b(t),x^i\big)$, leading to the following connection from Eq.(\ref{gaug}):
\begin{align}
    \hat{\Gamma}^0_{00}=\frac{\ddot{b}}{\dot{b}}=\gamma(t),\;\;\; \hat{\Gamma}^0_{0i}=\hat{\Gamma}^0_{i0}=\hat{\Gamma}^{0}_{ij}=\hat{\Gamma}^{a}_{ij}=0,
\end{align}
where the index $0$ represents the time coordinate and the indices $i$ and $j$ represent spacial coordinates. This choice of the vector $\xi^\mu$ can be unique if the spacetime invariance under rotations and translations is required \cite{NOJIRI2024}. In such a case, the nonmetricity takes the simpler form \cite{jimenez2020cosmology,koussour2022}:
%\begin{align}
 %   Q=6H^2,
    
%\end{align}
%where 'dot' ($\dot{}$) denotes differentiation w.r.t. cosmic time $t$ and Hubble rate is given by $H=\frac{\dot{a}}{a}$. Thus, for this spacetime metric, the nonmetricity scalar is given by:
\begin{align}
    Q=-6H^2,
    \label{eq18}
\end{align}
where $H(t)=\frac{\dot{a}}{a}$ is called Hubble function and dot ($\dot{}$) symbolises the derivative with respect to the time coordinate $t$ \cite{dodelson2020modern}.

Furthermore, we assume that the distribution of matter can be well represented by an isotropic perfect fluid characterised by the energy density $\rho$ and pressure $p$. The corresponding stress-energy tensor is:
\begin{align}
    T_{\mu\nu}=(\rho+p)u_\mu u_\nu+p g_{\mu\nu},
\label{eq15}
\end{align}
 where the four-velocity satisfies the normalisation condition $u^\mu u_\mu=-1$.
 
 %and \textcolor{red}{the matter Lagrangian reduces to} $\mathcal{L}_m = p$. 
%Hence, the auxiliary tensor $\Theta_{\alpha\beta}$ \textcolor{red}{takes the form} \cite{harko2018extensions,xu2019f}:
%\begin{align}
%    \Theta_{\mu\nu}=-2 T_{\mu\nu}+p g_{\mu\nu}.
%\label{eq16}
%\end{align}
In order to preserve homogeneity and isotropy, all physical quantities such as pressure, energy density, and scalar fields $\phi$ and $\psi$ are presumed to depend only on the time $t$. Moreover, the pressure and energy density are related via the equation of state (EoS):
\begin{align}
    p=w \rho,
\end{align}
characterized by EoS parameter $w$, which depends on the type of fluid. In this case, we assume that the cosmological fluid is composed of two different components, each represented by its own isotropic relativistic fluid. The first component is pressureless dust, which is described by an equation of state of the form $p_m=w_m \rho_m$ with $w_m=0$. The second component is radiation, chararacterized by an equation of state $p_r=w_r \rho_r$, where $w_r=1/3$. Thus, we can write the following total energy density and pressure of the cosmic fluid as:
\begin{align}
    \rho_{\rm tot} = \rho_r + \rho_m, \ \ \ \ p_{\rm tot}= \frac{1}{3}\rho_r,
    \label{eqfluids}
\end{align}
from which the resulting stress-energy tensor has the following components
\begin{align}
    T^\mu_{\nu}= \text{diag}\big(-\rho_r-\rho_m,\frac{1}{3}\rho_r,\frac{1}{3}\rho_r,\frac{1}{3}\rho_r\big),\label{eq:matter}
\end{align}
leading to the trace \begin{align}
    T=-\rho_m.
\end{align}
Now, let us turn our attention to the form of the matter Lagrangian $\mathcal{L}_m$. In fact, it was noted that the form of the on-shell perfect fluid matter Lagrangian $\mathcal{L}_m$ (for example $\mathcal{L}_m=p,\ \mathcal{L}_m=-\rho, \mathcal{L}_m=T$), influences the shape of the equations of motion \cite{bertolami2008nonminimal,avelino2018matter,ferreira2020lagrangian,avelino2022shell}. Among possible choices, recently it was argued that $\mathcal{L}_m = T$ should be used to describe baryons and dark matter \cite{ferreira2020lagrangian}. In fact, the consistency between the evolution of the energy and linear momentum of the particles requires the coincidence between $T$ and matter Lagrangian \cite{avelino2022shell,gonccalves2024dynamical}. Hence, in spirit of those works, we consider on-shell Lagrangian density $\mathcal{L}_m=T=-\rho_m$. Thus, assuming that $\mathcal{L}_m$ is linear with respect to the metric, the tensor $\Theta_{\mu\nu}$ takes the following form:
\begin{align}
    \Theta_{\mu\nu}=-2T_{\mu\nu}-\rho_m g_{\mu\nu},
\end{align}
which, upon using Eq. \eqref{eq:matter}, is found to be:
\begin{align}
    \Theta^\mu_\nu=\text{diag}\Big(\rho_m+2\rho_r,-\frac{2}{3}\rho_r-\rho_m,-\frac{2}{3}\rho_r-\rho_m,-\frac{2}{3}\rho_r-\rho_m\Big).
\end{align}

Under all of the assumptions outlined above for the geometry and matter distribution, the field equations of the scalar-tensor representation of $f(Q,T)$ in Eq. \eqref{eq10} is:
%\begin{align}
 %   \frac{1}{2}\Big[  \phi Q+\psi T -V  \Big]-6\phi H^2 = 8\pi \rho_{tot} +\psi(\rho_{tot}+p_{tot}),\\
  %    \frac{1}{2}\Big[  \phi Q+\psi T -V  \Big]-2(\phi \dot{H}+\dot{\phi}H)-6\phi H^2=- 8\pi p_{tot}
%\end{align}
\begin{align}
    3\phi H^2-\frac{V}{2}=8\pi(\rho_m+\rho_r)+\frac{1}{2}\psi(\rho_m+2\rho_r),
    \label{eqs1}
    \\
    -\frac{1}{2}V+2(\phi \dot{H}+\dot{\phi}H)+
    3\phi H^2=- \frac{8}{3}\pi \rho_{r}-\frac{1}{6}\psi(2\rho_r+3\rho_m).
    \label{eqs2}
\end{align}
%or minus before 3\phi H^2%

%$   \frac{1}{2}\Big[  \phi Q-V  \Big
%]-6\phi H^2 = 8\pi(\rho_m+\rho_r)+\frac{1}{2}\psi(\rho_m+2\rho_r)$ aaaa
%$\frac{1}{2}\Big[  \phi Q-V  \Big]-2(\phi \dot{H}+\dot{\phi}H)-6\phi H^2=- 8\pi p_{tot}+\frac{\psi}{6}(2\rho_r+\rho_m)$

%that can be combined to obtain:
%\begin{align}
% -\phi  \dot{H}-H \dot{\phi}-\frac{1}{2}\rho_m (\psi +8 \pi )-\frac{2}{3} \rho_r (\psi +8 \pi )=0,
% \label{hubble}
%\end{align}
%and will be useful in the formulation of the dynamical system. 
We note, that in the limit $\phi=1, \psi=0, \dot{\phi}=\dot{\psi}=0$, the Friedmann equations of standard Einstein's gravity are recovered \cite{jimenez2020cosmology}. On the other hand, the conservation equation in Eq. \eqref{eq:cons2} reads:
\begin{align}
    \frac{1}{2} \left(-\dot{\phi} V_\phi-\dot{\psi}  V_\psi+12 H
   \dot{H}\phi +6 H^2 \dot{\phi}-\psi \left(\dot{\rho}_m+2 \dot{\rho}_r\right)-16 \pi
    \left(\dot{\rho}_m+\dot{\rho}_r\right)-(\rho_m+2 \rho_r) \dot{\psi}\right)=0.
\end{align}

While the conservation of the matter-energy distribution is not essential within theories with geometry-matter couplings, it is reasonable to expect that our Universe does not destroy and create matter \cite{cipriano2024gravitationally}. Additionally, at late times, the conditions within Universe are not favourable for the possible transformations between different matter sectors. Thus, we can safely assume within our work that both dust and radiation are independently conserved. The conservation equations for dust and radiation take the forms:
\begin{align}
 \label{eqs3}
    \dot{\rho}_m+3H \rho_m=0,\\
     \dot{\rho}_r+4H \rho_r=0.
      \label{eqs4}
\end{align}
The conservation equations above can be used to simplify Eq. \eqref{eq:cons2} via the elimination of $\dot{\rho}_m$ and $\dot{\rho}_r$, thus resulting in a conservation equation for the field $\psi$:
\begin{align}
    \rho_r \dot{\psi}=-\frac{3}{2}H\rho_m \psi
     \label{eqs7}.
\end{align}
It is worth to notice that the same conservation law was obtained in scalar-tensor $f(R,T)$ gravity, where also $\mathcal{L}_m=T$ has been chosen \cite{gonccalves2024dynamical}. In fact, taking the limit $f(Q,T)\rightarrow Q+F(T)$ results in the same equations as in the scenario where $f(R,T) \rightarrow R+F(T)$, since the nonconservation of the stress-energy tensor is caused by the functional form of the coupling with the trace $T$ \cite{harko2018extensions}.

Finally, the equations of motion for the scalar fields $\phi$ and $\psi$ given in Eq. \eqref{eq12} take the forms:
\begin{align}
    V_\phi  \equiv\frac{\partial V}{\partial \phi}= 6H^2, \;\;\;\; V_\psi \equiv \frac{\partial V}{\partial \psi}=-\rho_m.
    \label{eqpotentials}
\end{align}
To simplify the analysis that follows, it is useful to rewrite the potential $V\left(\phi,\psi\right)$ as a function of time $t$. To do so, one can make use of the chain rule for the time derivatives of the potential $\dot{V}=V_\phi \dot{\phi}+V_\psi \dot{\psi}$ which, upon using Eq. \eqref{eqpotentials}, leads to the following equation that effectively replaces the equations of motion for the scalar fields:
\begin{align}
    \dot{V}=6 H^2 \dot{\phi} - \rho_m \dot{\psi}.
    \label{potred}
\end{align}
In this manner, instead of the two equations in Eq. (\ref{eqpotentials}), one can instead use Eq. \eqref{potred}, since it carries out the same information, while reducing number of the degrees of freedom. In fact, by considering $V$ as a function of fields $\psi$ and $\phi$, we deal with a potential that is expressed as a function of time $t$, reducing the number of degrees of freedom by one.

In summary, we have obtained a system of six equations, namely Eqs. \eqref{eqs1}, \eqref{eqs2}, \eqref{eqs3}, \eqref{eqs4}, \eqref{eqs7} and \eqref{potred}, for the five unknown functions $\rho_r$, $\rho_m$, $\phi$, $\psi$, and $V$. Note that only five of these equations are linearly independent, and thus the system is not over-determined. This can be proved by taking the derivative of Eq. \eqref{eqs1} and then using Eqs. \eqref{eqs2}, \eqref{eqs1}, \eqref{potred}, \eqref{eqs3}, \eqref{eqs4} to eliminate the terms $\dot\phi$, $\phi$, $\dot V$, $\dot\rho_m$, and $\dot\rho_r$, respectively, from which one recovers Eq. \eqref{eqs7}.

%%%%
\section{Dynamical system analysis of Scalar-tensor $f(Q,T)$ cosmology}\label{sec:dynsys}
%%%%
The dynamical systems approach is a useful tool for testing background cosmology \cite{bahamonde2018dynamical}. It allows for the identification and characterization of critical points associated with specific evolution behaviors, helping to unveil the nature of the dynamics, and was adapted in a broad range of the other modified gravity models \cite{gonccalves2024dynamical,rosa2024dynamical}. These critical points can then be correlated with the Universe's evolution as observed in cosmology, providing a strong test for the scalar-tensor $f(Q,T)$ gravity introduced herein. For more details on the dynamical system approach to cosmology, we refer the reader to the review \cite{bahamonde2018dynamical}.

In what follows, in order to examine the equations outlined in the previous section using the dynamical system approach, one has to define a new set of dimensionless dynamical variables. For the purpose of this work, we introduce one dynamical variable for each of the five unknown quantities in the system, namely:
\begin{align}
    \Phi=\phi,\ \ \ \Psi=\frac{\psi}{8 \pi}, \ \ \  \Omega_m = \frac{8 \pi \rho_m}{3H^2}, \ \ \ \ \Omega_r = \frac{8 \pi \rho_r}{3H^2}, \ \ \ U=\frac{V}{6H^2}
    \label{variables}
\end{align}
Moreover, it is useful to introduce the deceleration parameter \cite{dodelson2020modern}:
\begin{align}
    \mathcal{Q}=-\frac{\ddot{a}}{aH^2}=-(1+\frac{\dot H}{H^2}), 
    \label{decceleration}
\end{align}
as it is directly associated with the dynamical behaviour of the scale factor $a$. Importantly, cosmic deceleration occurs for $\mathcal Q>0$, while accelerated phase corresponds to the $\mathcal Q<0$.

The study of the evolution of a dynamical system requires the introduction of a dimensionless time parameter. In this work, we use the number of e-folds, $N\equiv \ln(a /a_0)$, where  $a_0$ corresponds to the present value of the scale factor, as the dimensionless time parameter \cite{dodelson2020modern}. Then, the equations can be rewritten in terms of $N$ with the aid of the following chain rule for the time derivatives:
\begin{align}
X'\equiv  \frac{d X}{d N}= \frac{\dot X}{H}.
\label{efoldingderivative}
\end{align}

\noindent Using the definitions introduced above, the field equations in Eqs. \eqref{eqs1} and \eqref{eqs2}, take the forms:
\begin{equation}\label{con1}
    \Phi=\Omega_m\left(1+\frac{\Psi}{2}\right)+\Omega_r\left(1+\Psi\right)+U,
\end{equation}
\begin{equation}\label{con2}
    \Phi\left(1-2\mathcal Q\right)+2\Phi'=3U-\Omega_r\left(1+\Psi\right)-\frac{3}{2}\Psi\Omega_m,
\end{equation}
respectively. Furthermore, the system of equations \eqref{eqs3}, \eqref{eqs4}, \eqref{eqs7}, and \eqref{potred} become, under the same definitions:
\begin{align}
\Omega'_r=2(\mathcal{Q}-1)\Omega_r,\label{dyn1} \\ \Omega'_m=(2\mathcal{Q}-1)\Omega_m, \label{dyn2}\\
   \Psi'\Omega_r=-\frac{3}{2}\Psi \Omega_m, \label{dyn3}\\
     U'= 2 (\mathcal{Q}+1) U-\frac{1}{2}\Omega_m \Psi '+\Phi' \label{dyn4}.
\end{align}
These equations serve as dynamical equations for the dynamical variables $\{\Psi,U,\Omega_m,\Omega_r\}$.

From Eqs. \eqref{con1} and \eqref{con2} one can observe that the cosmological equations in GR are obtained in the limit $\Phi=1$ and $\Psi=0$. In the general $f(Q,T)$ case, Eq. \eqref{con1} serves as a constraint that allows one to eliminate one of the dynamical variables from the system, whereas Eq. \eqref{con2} serves as a dynamical equation for $\Phi$ due to its dependency on $\Phi'$. The dynamical system in Eqs. \eqref{con2} to \eqref{dyn4} present a total of four fixed points which are summarised in Table \ref{tab:fixed}. These fixed points are analogous to those of GR, i.e., one obtains set of fixed points corresponding to a radiation domination era $\mathcal A$ for arbitrary values of $\Phi$ and $\Psi$ with $\mathcal Q=1$, a set of fixed points corresponding to a matter domination era $\mathcal B$ for $\Psi=0$ and arbitrary $\Phi$ with $\mathcal Q=\frac{1}{2}$, a set of fixed points corresponding to late-time cosmic acceleration $\mathcal C$ for arbitrary values of $\Phi$ and $\Psi$ with $\mathcal Q=-1$, and a set of fixed points corresponding to vacuum universes for arbitrary values of $\Psi$ with arbitrary $\mathcal Q$, the latter being the least interesting from the point of view of cosmological evolution in our universe. The stability of these fixed points can be extracted from a typical analysis of eigenvalues (see e.g. \cite{Perko2001,Wiggins2003}), from which one verifies that $\mathcal A$ are reppelers, $\mathcal B$ are saddle points, and $\mathcal C$ are attractors in the phase space, a behaviour qualitatively similar to that of GR and allowing for a cosmological evolution starting from a radiation dominated phase, transitioning to a matter dominated phase, and transitioning later to a late-time cosmic accelerated period.  

\begin{table}
    \centering
    \begin{tabular}{ c | c | c | c | c | c | c }
         &  $\Omega_r$ & $\Omega_m$ & $U$ & $\Phi$ & $\Psi$ & $\mathcal Q$ \\ \hline
        $\mathcal A$ & $\frac{\Phi}{1+\Psi}$ & $0$ & $0$ & ind. & ind. & $1$ \\
        $\mathcal B$ & $0$ & $\Phi$ & $0$ & ind. & $0$ & $\frac{1}{2}$ \\
        $\mathcal C$ & $0$ & $0$ & $\Phi$ & ind. & ind. & $-1$ \\
        $\mathcal D$ & $0$ & $0$ & $0$ & $0$ & ind. & ind. \\ 
    \end{tabular}
    \caption{Fixed points of the dynamical system given by Eqs. \eqref{con2} to \eqref{dyn4}. The tag "ind." indicates that the value of this variable is arbitrary, i.e., the fixed points are not isolated but rather one- or two-dimensional sub-manifolds.}
    \label{tab:fixed}
\end{table}

We are interested in verifying if the dynamical system given in Eqs. \eqref{con2} to \eqref{dyn4} allows for cosmological solutions qualitatively similar to those of GR. To simplify this task, instead of numerically solving the entire system simultaneously, we implement a reconstruction method that allows us to obtain separately the solutions for the density parameters $\Omega_r$ and $\Omega_m$, and the deceleration parameter $\mathcal Q$, in the GR limit, and then use these results to obtain the solutions for the scalar fields $\Phi$ and $\Psi$, and the scalar potential $U$, compatible with the GR limit.

\begin{figure*}
    \includegraphics[scale=0.54]{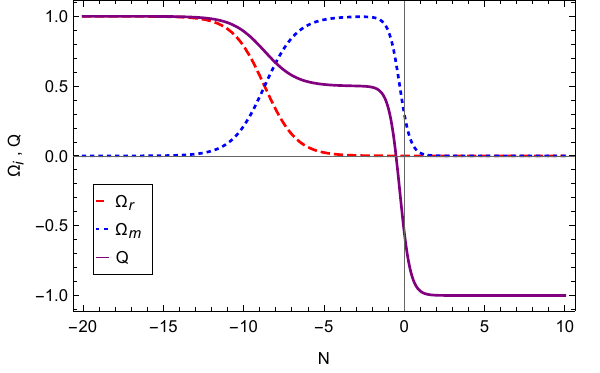}\qquad
    \includegraphics[scale=0.54]{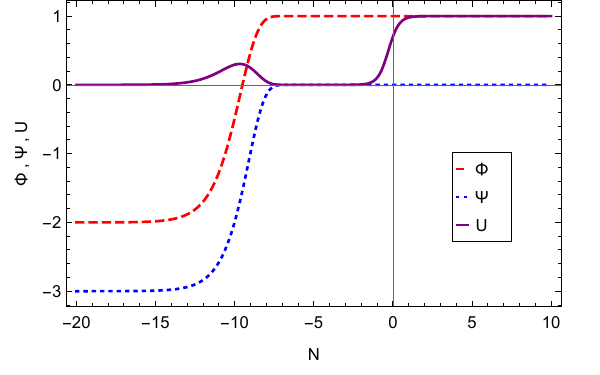}\qquad
    \includegraphics[scale=0.54]{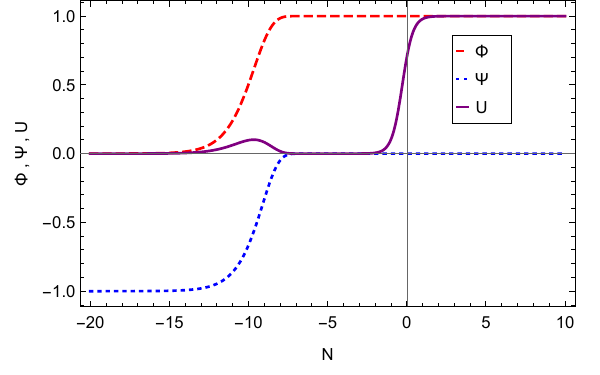}\\
    \includegraphics[scale=0.54]{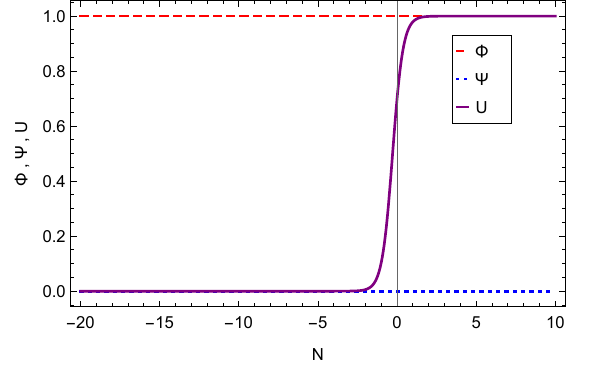}\qquad
    \includegraphics[scale=0.54]{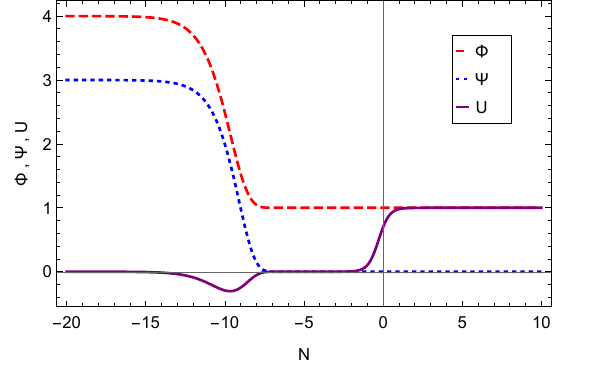}\qquad
    \includegraphics[scale=0.54]{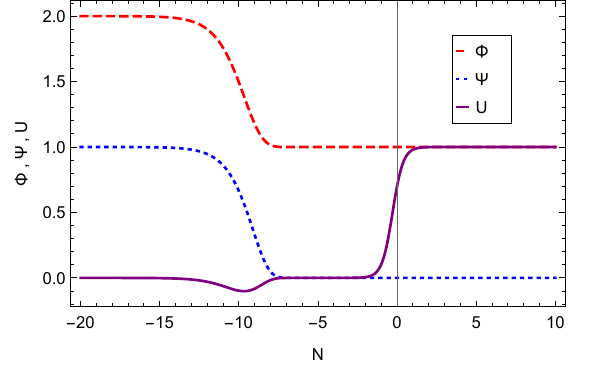}
    \caption{Density parameters $\Omega_r$ and $\Omega_m$ and deceleration parameter $\mathcal Q$ as a function of $N$ (top left panel), and scalar fields $\Phi$ and $\Psi$, and scalar potential $U$ as a function of $N$ in the GR limit corresponding to $\Psi(-20)=0$ (bottom left panel) and for the initial conditions $\Psi_0=-3$ (top middle panel), $\Psi_0=-1$ (top right panel), $\Psi_0=3$ (bottom middle panel), and $\Psi_0=1$ (bottom right panel).}
    \label{fig:parameters}
\end{figure*}

We start by taking the limit to GR, i.e., $\Phi=1$ and $\Psi=0$, into Eqs. \eqref{con1} and \eqref{con2}. We solve the resulting Eq. \eqref{con1} with respect to the scalar potential $U$ and use the result to eliminate this quantity from the resulting Eq. \eqref{con2}. The system of the resulting Eqs. \eqref{con2} to \eqref{dyn2} is then numerically integrated, subjected to the set of initial conditions $\Omega_r(0)=5 \times 10^{-5}$, $\Omega_m(0)=0.3$, and $\mathcal Q(0)=-0.55$, compatible with the observations from the Planck satellite \cite{Planck:2018vyg}, from which one obtains the solutions for $\Omega_r$, $\Omega_m$, and $\mathcal Q$ in the GR limit. These solutions are plotted in the top left panel of Fig. \ref{fig:parameters}. As expected, we observe that the evolution is dominated by radiation at early times, with $\Omega_r\simeq 1$ and $\mathcal Q\simeq 1$, followed by a transition to a matter dominated phase with $\Omega_m\simeq 1$ and $\mathcal Q\simeq \frac{1}{2}$, and finally transitioning to a late-time cosmic acceleration period with $\Omega_r\simeq0$, $\Omega_m\simeq 0$, and $\mathcal Q\simeq -1$.

The solutions obtained can be reinserted back into Eqs. \eqref{dyn4}, \eqref{dyn3}, and \eqref{con1} to obtain the solutions for $\Psi$, $U$, and $\Phi$, in this specific order, and under a particular choice of initial condition for $\Psi$. This reconstruction method allows us to extract the solutions for $\Phi$, $\Psi$, and $U$ compatible with the $\Lambda$CDM model. To guarantee that the solutions obtained are not only compatible with current cosmological observations, but also compatible with the observed solar system dynamics in the weak field regime, it is strictly necessary that, at present times $N=0$, the scalar fields satisfy the initial conditions $\Psi(0)\simeq 0$ and $\Phi(0)\simeq 1$. The first of these conditions can be shown to be always satisfied due to the exponentially decaying behaviour of the scalar field $\Psi$ during the matter dominated phase. This can be observed by a direct integration of Eq. \eqref{dyn4}, which results in:
\begin{equation}
    \Psi(N)=C\exp\left(\int_1^{N}-\frac{3\Omega_m(y)}{2\Omega_r(y)}dy\right),
\end{equation}
for some constant of integration $C$. This result implies that during the radiation dominated phase, where $\Omega_r\simeq 1$ and $\Omega_m\simeq 0$, the scalar field $\Psi$ is approximately constant. However, in the transition to the matter dominated phase where $\Omega_r\simeq 0$ and $\Omega_m\simeq 1$, the scalar field $\Psi$ decays exponentially fast, with the rate of decay increasing during the transition. As a result, the effects of the scalar field $\Psi$ are exponentially suppressed during the matter domination phase and the initial condition $\Psi(0)\simeq 0$ is always satisfied. Following this result and the fact that the solutions for $\Omega_r$ and $\Omega_r$ were obtained in the GR limit, Eq. \eqref{con1} implies that the second initial condition $\Phi(0)\simeq 1$ is always satisfied as long as the initial condition for $U$ is $U(0)=0.69995$, i.e., the scalar potential $U$ effectively plays the role of the cosmological constant at present times. 

\begin{figure*}
    \includegraphics[scale=0.54]{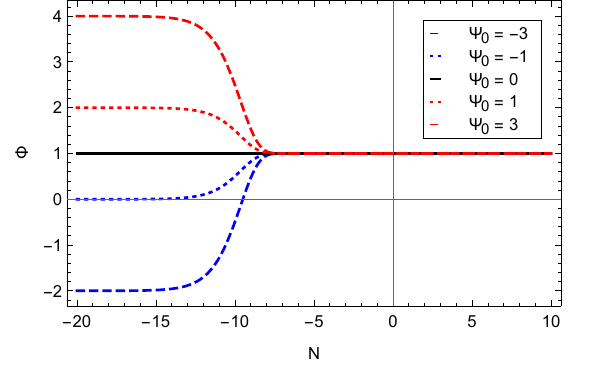}\qquad
    \includegraphics[scale=0.54]{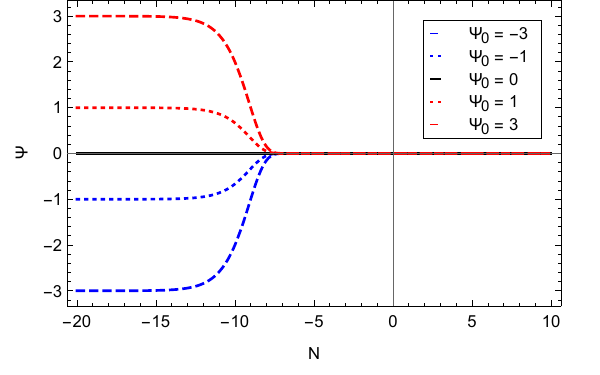}\qquad
    \includegraphics[scale=0.54]{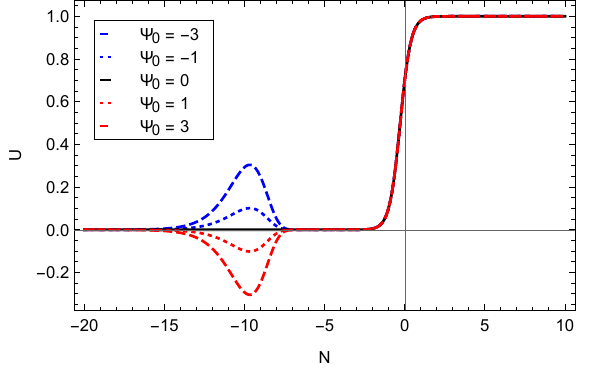}
    \caption{Scalar fields $\Phi$ (left panel) and $\Psi$ (middle panel), and scalar potential $U$ (right panel) as a function of $N$ for varying values of the initial condition $\Psi_0=\{-3,-1,0,1,3\}$.}
    \label{fig:initials}
\end{figure*}

Following the analysis outlined above, the only free parameter in the presented cosmological solutions is the value of the scalar field $\Psi$ at early times, before being exponentially suppressed. For the purpose of this work, we take the value $N=-20$ as our earliest instant in time, given that at this time the system is still in the radiation dominated era for more than $5$ e-folds. If one takes the initial condition $\Psi(-20)\equiv\Psi_0=0$, the GR limit is recovered, where $\Phi=1$ and $\Psi=0$, and $U$ plays the role of the cosmological constant in the $\Lambda$CDM model (see the bottom left plot in Fig. \ref{fig:parameters}). By changing this initial condition, the solutions for $\Phi$, $\Psi$, and $U$ are perturbed, while $\Omega_r$, $\Omega_m$, and $\mathcal Q$ remain the same. For the initial conditions $\Psi_0=\{-3, -1, 1, 3\}$, the solutions for $\Phi$, $\Psi$, and $U$ can be found in the middle and right panels of Fig. \ref{fig:parameters}. Furthermore, the effects of this initial condition on the solutions for $\Phi$, $\Psi$, and $U$ are also clarified in Fig. \ref{fig:initials}.

Our results in Fig. \ref{fig:parameters} indicate that, for any initial condition different from the GR limit $\Psi_0=0$, the scalar field $\Psi$ is approximately constant at early times and decays exponentially during the matter dominated phase, as expected. The scalar field $\Phi$ presents a qualitatively similar behaviour but asymptotically approaches the value $\Phi=1$. The scalar potential $U$ is approximately zero at early times and, during the transition between the radiation and matter dominated phases, presents a kink that lasts for approximately $5$ e-folds. It is during this kink that the scalar fields $\Phi$ and $\Psi$ suffer their exponential decay, with a positive kink associated with negative scalar fields and vice versa. At late times, the scalar potential $U$ grows asymptotically to $1$ during the late-time cosmic acceleration period. From Fig. \ref{fig:initials} one observes that larger absolute values of $\Psi_0$, which correspond to larger amplitudes of decay for the scalar fields $\Phi$ and $\Psi$, induce larger kinks in the potential $U$, with opposite signs. It thus seems that the effects of $f\left(Q,T\right)$ gravity are particularly important at early times, when the scalar fields deviate the most from their GR counterparts, and at late times, when the potential $U$ takes the place of the cosmological constant. 

\section{Summary and Conclusions}\label{sec:concl}

In this work, we have introduced an alternative dynamically equivalent scalar-tensor representation for the $f(Q,T)$ gravity and studied its dynamical behaviour for the FLRW cosmology. In this representation, the $f(Q,T)$ functional from Eq. \eqref{eq7} has been reformulated using two fields $\phi,\psi$ taking the form of Eq. \eqref{eq10}. The scalar–tensor representation of the theory not only greatly simplifies the analysis compared to previous studies that used the geometrical representation but also allows one to perform the analysis in a model-independent way, without the requirement of selecting specific forms of the function \cite{narawade2023}. In this framework, the standard density parameters of the $\Lambda$CDM model from GR can be used, enabling a direct comparison of the model with observational data. We have modelled the spacetime geometry using the FLRW metric, with the matter distribution represented by the two independently conserved relativistic perfect fluids: one acting as dust-like matter ($\rho_m, p_m$) and the other as radiation ($\rho_r, p_r$). Based on this choice of the cosmological fluids, we adopted the on-shell matter Lagrangian $\mathcal{L}_m=T=-\rho_m$, in accordance with the recent literature \cite{avelino2018matter,ferreira2020lagrangian,avelino2022shell,gonccalves2024dynamical}. Additionally, we have adapted a gauge for connection, that is consistent with the FLRW  spacetime \cite{NOJIRI2024}.

As a first step, we have analysed the limiting case where $\Phi = 1$ and $\Psi = 0$, where we have shown that the system reduces to the standard cosmological equations of General Relativity (GR). However, in more general $f(Q,T)$ framework, the system reveals richer structure with several distinct solutions. The dynamical system derived from these equations identifies four sets of fixed points, each set being associated with one of the fixed points in standard GR: radiation domination ($\mathcal{A}$), matter domination ($\mathcal{B}$), late-time cosmic acceleration ($\mathcal{C}$), and vacuum solutions with open geometry ($\mathcal{D}$). A stability analysis based on eigenvalue examination have shown that radiation-dominated fixed points are unstable, matter-dominated fixed points are saddle points, and late-time cosmic acceleration fixed points are stable. The obtained structure of the phase space closely resembles that of standard GR \cite{dodelson2020modern}, indicating that the scalar-tensor $f(Q,T)$ gravity is capable of producing cosmological solutions for which the cosmic evolution transitions from a radiation-dominated era to a matter-domination era, and then to a late-time acceleration. 

To prove that the solutions described above exist, we have followed a reconstruction method in which we have obtained the solutions for the density parameters $\Omega_r$ and $\Omega_m$, as well as the deceleration parameter $\mathcal Q$ in the GR limit, and we reintroduced those solutions into the general equations of scalar-tensor $f\left(Q,T\right)$ gravity to extract the solutions for the scalar fields $\Phi$ and $\Psi$ and the scalar potential $U$. The results show a transition from a radiation-dominated era, where $\Omega_r \approx 1$ and $Q \approx 1$, to a matter-dominated phase with $\Omega_m \approx 1$ and $Q \approx \frac{1}{2}$, and finally to a late-time accelerated expansion where both $\Omega_r$ and $\Omega_m$ approach zero and $Q \approx -1$. In this way, a cosmological evolution qualitatively similar to that of the $\Lambda$CDM model was obtained. Our results indicate that the scalar fields are exponentially suppressed during the matter-dominated phase, and thus they play no role at present time and the future dynamics, leading to a consistency of the model with the weak-field solar system dynamics. On the other hand, the scalar potential $U$ is negligible at early times, kinks shortly during the time period when the scalar fields suffer a decay, and finally increases at the end of the matter domination era, thus playing the role of the cosmological constant $\Lambda$.

In summary, our findings show that while scalar-tensor $f(Q,T)$ gravity retains the key aspects of cosmological evolution found in GR, the additional degrees of freedom in the model provide greater flexibility. Moreover, late-time $\Lambda$CDM dynamics have been obtained without referring to the Dark Sector \cite{arbey2021}. In this manner, the $f(Q,T)$ theory in the scalar-tensor form can successfully reproduce a meaningful and physically relevant cosmological background evolution. However, the fact that the cosmological evolution is virtually indistinguishable from GR renders this theory unfalsifiable from a cosmological point of view alone. A more comprehensive analysis of perturbations would be crucial to establish the viability of the scalar-tensor $f(Q,T)$ gravity. Furthermore, to gain more comprehensive understanding of the role played by the scalar-tensor $f(Q,T)$ gravity in cosmology, our upcoming investigations will be focused on comparing this theory with a current observational data. In particular, our next phase of research aims to constrain the theory using various datasets \cite{arora2020f}. Thus, important cosmological and astrophysical tests remain to be studied.

\bibliographystyle{apsrev}
\bibliography{bibliography}% Produces the bibliography via BibTeX.

\end{document}